\documentclass[12pt]{article}
\usepackage{bbold,hyperref,graphicx,color}

\newcommand{\be}{\begin{equation}}
\newcommand{\ee}{\end{equation}}
\newcommand{\bea}{\begin{eqnarray}}
\newcommand{\eea}{\end{eqnarray}}
\newcommand{\beas}{\begin{eqnarray*}}
\newcommand{\eeas}{\end{eqnarray*}}

\begin{document}
\begin{titlepage}

\begin{flushright}
{\small OU-HET-1118}
\end{flushright}

\medskip

\begin{center}

{\Large Defining entanglement without tensor factoring: a Euclidean hourglass prescription}

\vspace{12mm}

\renewcommand\thefootnote{\mbox{$\fnsymbol{footnote}$}}
Takanori Anegawa${}^{1}$\footnote{takanegawa@gmail.com},
Norihiro Iizuka${}^{1}$\footnote{iizuka@phys.sci.osaka-u.ac.jp},
Daniel Kabat${}^{2,3}$\footnote{daniel.kabat@lehman.cuny.edu}

\vspace{6mm}

${}^1${\small \sl Department of Physics, Osaka University} \\
{\small \sl Toyonaka, Osaka 560-0043, JAPAN}

\vspace{3mm}

${}^2${\small \sl Department of Physics and Astronomy} \\
{\small \sl Lehman College, City University of New York} \\
{\small \sl 250 Bedford Park Blvd.\ W, Bronx NY 10468, USA}

\vspace{3mm}

${}^3${\small \sl Graduate School and University Center, City University of New York} \\
{\small \sl  365 Fifth Avenue, New York NY 10016, USA}

\end{center}

\vspace{12mm}

\noindent
We consider entanglement across a planar boundary in flat space.  Entanglement entropy is usually thought of as the von Neumann entropy of a reduced density matrix,
but it can also be thought of as half the von Neumann entropy of a product of reduced density matrices on the left and right.  The latter form allows a natural
regulator in which two cones are smoothed into a Euclidean hourglass geometry.  Since there is no need to tensor-factor the Hilbert space, the regulated entropy is
manifestly gauge-invariant and has a manifest state-counting interpretation.  We explore this prescription for scalar fields, where the entropy is insensitive to
a non-minimal coupling, and for Maxwell fields, which have the same entropy as $d-2$ scalars.

\end{titlepage}
\setcounter{footnote}{0}
\renewcommand\thefootnote{\mbox{\arabic{footnote}}}

\hrule
\tableofcontents
\bigskip
\hrule

\addtolength{\parskip}{8pt}
%%%%%%%%%%%%%%%%%%%%%%%%%%
\section{Introduction}
%%%%%%%%%%%%%%%%%%%%%%%%%%
Consider a field theory in its Minkowski ground state and divide space into $L \cup R$ where
\be
L = \lbrace x < 0 \rbrace \qquad R = \lbrace x > 0 \rbrace
\ee
Entanglement entropy is usually defined as the von Neumann entropy of the reduced density matrix associated with the region $R$ (or equivalently the region $L$).
\bea
\nonumber
&& \rho_R = {\rm Tr}_L \big( \vert 0 \rangle \langle 0 \vert \big) \\[2pt]
\label{SR}
&& S = - {\rm Tr}_R \big(\rho_R \log \rho_R\big)
\eea
But this definition is problematic in a continuum field theory for a number of reasons.  For one thing the Hilbert space of a field theory does not factorize
into ${\cal H}_L \otimes {\cal H}_R$, so it's not clear that the entropy (\ref{SR}) is well-defined or can be thought of as counting states.  This is closely related to the fact
that for a field theory entanglement entropy is UV divergent and requires regularization.  For a recent discussion of these matters see \cite{Witten:2018lha}.
One way to address UV issues would be to define the field theory on a lattice.  But even this isn't fully satisfactory for a gauge theory, where the connection variables
live on links and the decomposition into $L \cup R$ isn't gauge invariant.  For work on this issue see \cite{Donnelly:2011hn,Eling:2013aqa,Casini:2013rba,Casini:2014aia,Donnelly:2014fua,Huang:2014pfa,Ghosh:2015iwa,Aoki:2015bsa,Donnelly:2015hxa,Soni:2015yga,Casini:2015dsg,Soni:2016ogt,Agarwal:2016cir,Aoki:2017ntc,Huerta:2018xvl,Casini:2019nmu}.

It would be desirable to have a regulated definition of entanglement entropy in field theory that is manifestly gauge invariant and has a manifest state-counting interpretation.
Here we make a simple proposal in this direction.  For motivation consider a field theory in $1+1$ dimensions with coordinates $(t,x)$.  On the $t = 0$ time slice a Lorentz boost
is generated by
\be
K = \int_{-\infty}^\infty dx \, x T_{00}
\ee
This can be formally decomposed as $K = K_R - K_L$ where
\beas
&& K_R = \int_0^\infty dx \, x T_{00} \\
&& K_L = \int_{-\infty}^0 dx \, (-x) T_{00}
\eeas
are defined to boost the right and left half-spaces forward in time.  For a field theory in its ground state the modular operator is given by \cite{Witten:2018lha,Bisognano:1976za}
\be
\Delta = e^{-2 \pi K} = e^{-2 \pi (K_R - K_L)}
\ee
The modular operator can be formally decomposed as
\be
\Delta = \rho_L^{-1} \otimes \rho_R
\ee
where $\rho_L$, $\rho_R$ are the reduced density matrices describing the left and right half-spaces.  The modular operator is well-defined in the continuum but it's not directly
useful for studying entanglement.  Instead to study entanglement we'd like to work with the operator
\be
\label{Vdef}
V = K_L + K_R = \int_{-\infty}^\infty dx \, \vert x \vert T_{00}
\ee
which boosts both the left and right half-spaces forward in time.  Then formally we'd have
\be
e^{-2 \pi V} = \rho_L \otimes \rho_R
\ee
If we could define a partition function
\be
\label{Z}
Z(\beta) = {\rm Tr} \, e^{-\beta V} = {\rm Tr} \, \left( \rho_L^{\beta/2\pi} \otimes \rho_R^{\beta/2\pi}\right)
\ee
we could compute entanglement entropy using
\be
\label{S}
S_{\rm entanglement} = {1 \over 2} \left.\left(\beta {\partial \over \partial \beta} - 1\right)\right\vert_{\beta = 2 \pi} \big(- \log Z(\beta)\big)
\ee
The partition function (\ref{Z}) can be thought of as putting the field theory on two Euclidean cones of angle $\beta$ joined at their tips.
The factor of $1/2$ in (\ref{S}) compensates for the double-counting of having two cones and gives an entropy that formally agrees with (\ref{SR}).

\begin{figure}
\centerline{\includegraphics[width=7cm]{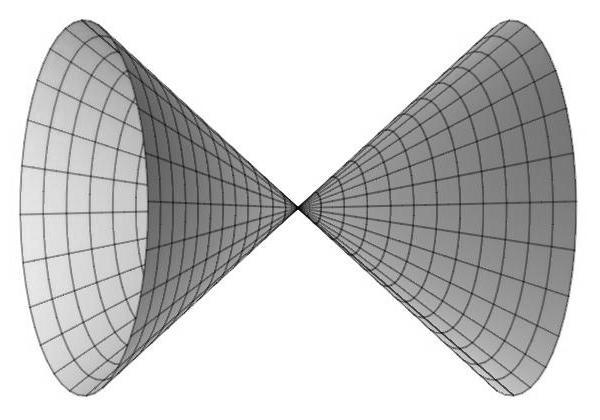} \hspace{2mm} \raisebox{2.3cm}{$\Rightarrow$} \hspace{2mm} \includegraphics[width=7cm]{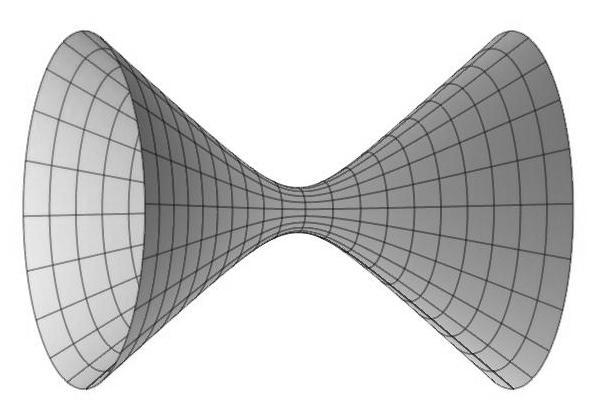}}
\caption{A double cone, smoothed near the origin into an hourglass geometry.\label{fig:hourglass}}
\end{figure}

The operator $V$ has UV problems near $x = 0$, but this is easy to regulate.  Introduce a small parameter $\epsilon$ and
choose a function $r_\epsilon(x)$ with
\be
r_\epsilon(x) \approx \left\lbrace\begin{array}{l}
\hbox{$\vert x \vert$ for $\vert x \vert > \epsilon$} \\
\hbox{\rm smooth and non-zero for $\vert x \vert < \epsilon$}
\end{array}\right.
\ee
In our explicit calculations we will generally take $r_\epsilon(x) = \sqrt{x^2 + \epsilon^2}$, but other choices are possible.
Given a choice of $r_\epsilon(x)$ we can define
\bea
\nonumber
&& V_\epsilon = \int_{-\infty}^\infty dx \, r_\epsilon(x) T_{00} \\
&& Z_\epsilon(\beta) = {\rm Tr} \, e^{-\beta V_\epsilon}
\eea
Note that the trace here is over the full Hilbert space, with no need to factor into ${\cal H}_L \otimes {\cal H}_R$.
The regulated partition function $Z_\epsilon(\beta)$ can be used to compute a regulated entanglement entropy
\be
\label{Sepsilon}
S_\epsilon = {1 \over 2} \left.\left(\beta {\partial \over \partial \beta} - 1\right)\right\vert_{\beta = 2 \pi} \big(- \log Z_\epsilon(\beta)\big)
\ee
This has a manifest state-counting interpretation, as (half of) the von Neumann entropy of the density matrix
\be
\label{rho-epsilon}
\rho_\epsilon = {1 \over Z_\epsilon(\beta)} e^{-\beta V_\epsilon}
\ee
Note that this prescription is manifestly gauge invariant and avoids any discussion of factoring the Hilbert space.  The prescription
is somewhat arbitrary, of course.  But in field theory some regulator is necessary if entropy of entanglement is to be discussed at all,
and the regulator we have introduced has the appealing features we just described.
Geometrically the regulated partition function has the interpretation of putting the field theory on a Euclidean spacetime with metric
\be
ds^2 = dx^2 + \big(r_\epsilon(x)\big)^2 d\theta^2 \qquad -\infty < x < \infty,\quad \theta \approx \theta + \beta
\ee
The regulator smooths the two cones into an hourglass shape as shown in Fig.\ \ref{fig:hourglass}.\footnote{This regulator was first introduced by Solodukhin, who applied it to the thermal atmosphere of a BTZ black hole \cite{Solodukhin:2004rv,Solodukhin:2005qy}.}  The hourglass geometry corresponds to studying the
theory with a position-dependent proper temperature
\be
\label{Tproper}
T_{\rm proper}(x) = {1 \over \beta r_\epsilon(x)}
\ee
The proper temperature is high near the waist of the hourglass and (as in Rindler space) falls off at large $\vert x \vert$, where $T_{\rm proper} \sim 1 / 2 \pi \vert x \vert$ when $\beta = 2 \pi$.

We will work with regulators $r_\epsilon(x)$ that are smooth near $x = 0$, so that we avoid UV problems in defining the operator $V_\epsilon$ or equivalently so that the hourglass geometry is
smooth.  This may not be a necessary requirement.  One is free to consider non-smooth regulators such as $r_\epsilon(x) = \vert x \vert + \epsilon$.  Such regulators
introduce curvature singularities on the hourglass geometry, which presumably make non-minimal couplings such as $R^2 \phi^2$ ill-defined, but if such problems can be dealt with or
avoided then non-smooth regulators are allowed and may provide a convenient calculational tool.

It's worth contrasting the hourglass prescription with a well-known approach in the literature, where entanglement entropy is defined by putting the field theory on a cone and varying with respect to the deficit angle \cite{Susskind:1993ws}.  The cone approach can be thought of as directly attempting to make sense of (\ref{SR}).  The cone partition function is UV divergent but can be regulated in a gauge-invariant manner.  However the resulting entropy, defined as $S_{\rm cone} = (\beta \partial_\beta - 1)(- \log Z_{\rm cone})$, can't necessarily be given a state-counting
interpretation.  This is particularly clear for scalar fields with non-minimal couplings to curvature.  Such fields respond to the curvature at the tip of the cone, even though the
non-minimal coupling doesn't affect the ground-state wavefunction in flat space and therefore shouldn't affect entanglement.  A closely-related example is a Maxwell field which can
lead to a negative entropy when put on a cone \cite{Kabat:1995eq,Donnelly:2012st}.

In what follows we illustrate these properties in a series of examples.  To warm up we begin by studying a general CFT and a free massive scalar in two dimensions,
reproducing standard results.  Then we consider non-minimal scalars and Maxwell fields in two dimensions, where the hourglass prescription makes an important difference: it leads to an
entropy which is independent of $\xi$ (for non-minimal scalars) and which vanishes (for Maxwell fields in two dimensions).  We consider scalar theories in higher dimensions in section \ref{sect:higher},
treat Maxwell fields in higher dimensions in section \ref{sect:higherMaxwell}, and conclude in section \ref{sect:conclusions}.

%%%%%%%%%%%%%%%%%%%%%%%%%%
\section{2D CFT\label{sect:CFT}}
%%%%%%%%%%%%%%%%%%%%%%%%%%
To get started we show that the hourglass prescription reproduces the standard results for entanglement in a 2D CFT \cite{Calabrese:2004eu,Holzhey:1994we}.

For a discussion that runs parallel to what follows, we consider a CFT on a spatial interval $0 < \sigma < \pi$.  The finite interval provides an infrared cutoff.
The Virasoro generators are given by
\be
L_m = \int_0^\pi d\sigma \, (\cos(m\sigma) T_{00} + i \sin(m\sigma) T_{01})
\ee
The vacuum state is annihilated by the $SL(2,{\mathbb R})$ algebra generated by $L_1$, $L_0$, $L_{-1}$.
We divide space in half, with $L = (0,\pi/2)$ and $R = (\pi/2,\pi)$.  The modular Hamiltonian -- the operator which annihilates the vacuum, leaves the midpoint fixed, and looks locally
like a boost near the midpoint -- is given by \cite{Hislop:1981uh,Casini:2011kv}\footnote{More generally if we divided space at $\sigma = a$ we could have first performed an $SL(2,{\mathbb R})$ transformation to move $a$ to $\pi/2$.}
\be
K = \int_0^\pi d\sigma \, (- \cos \sigma) T_{00} = - {1 \over 2} \left(L_1 + L_{-1}\right)
\ee
The operator $V$ is then given by
\be
V = \int_0^\pi d\sigma \, \vert \cos \sigma \vert \, T_{00}
\ee
We can regulate this by defining
\be
V_\epsilon = \int_0^\pi d\sigma \, r_\epsilon(\sigma) T_{00}
\ee
The exact form of $r_\epsilon(\sigma)$ won't matter, but an explicit example to keep in mind is
\be
\label{reps}
r_\epsilon(\sigma) = \sqrt{\cos^2 \sigma + \epsilon^2}
\ee
The partition function $Z_\epsilon(\beta) = {\rm Tr} \, e^{-\beta V_\epsilon}$ corresponds to putting the theory on a Euclidean spacetime with metric
\be
ds^2 = d\sigma^2 + \big(r_\epsilon(\sigma)\big)^2 d\theta^2 \qquad 0 < \sigma < \pi, \quad \theta \approx \theta + \beta
\ee
Defining
\be
\label{y}
y = \int_0^\sigma {d\sigma' \over r_\epsilon(\sigma')}
\ee
the metric becomes conformally flat, with
\be
\label{cylinder}
ds^2 = \big(r_\epsilon(\sigma)\big)^2 \big(dy^2 + d\theta^2\big)
\ee
We can drop the conformal factor to obtain a CFT on a cylinder, but the length of the spatial interval has changed from $\pi$ to
\be
L_\epsilon = \int_0^\pi {d\sigma' \over r_\epsilon(\sigma')}
\ee
For any reasonable regulator $r_\epsilon(\sigma)$ the length diverges logarithmically as $\epsilon \rightarrow 0$, $L_\epsilon \sim 2 \log (1 / \epsilon)$.  In particular this is true for the choice (\ref{reps}).
Let us pause to see how this comes about.  The divergence arises from behavior near the entangling surface $\sigma = \pi/2$ where
\be
r_\epsilon(\sigma) \approx \sqrt{\Big(\sigma - {\pi \over 2}\Big)^2 + \epsilon^2}
\ee
The divergence in $L_\epsilon$ arises from the integration region $\epsilon \ll \vert \sigma - {\pi \over 2} \vert \ll 1$ where
$r_\epsilon(\sigma) \approx \vert \sigma - {\pi \over 2} \vert$.  The contribution of this region is
\be
L_\epsilon \approx 2 \int\limits_{{\pi \over 2} + \epsilon} {d\sigma \over \sigma - {\pi \over 2}} \sim 2 \log (1/\epsilon)
\ee
Any reasonable regulator will lead to the same result, since the only role of the regulator in this calculation
is to provide a cutoff near the entangling surface.

On a cylinder we have thermal entropy given by\footnote{This follows from the Cardy formula, or more
directly from the result that in the limit of a long cylinder the partition function per unit length is $- \log Z / L_\epsilon = - {\pi c \over 6 \beta}$ \cite{Bloete:1986qm}.}
\be
S = {\pi c L_\epsilon \over 3 \beta}
\ee
Setting $\beta = 2\pi$ and dividing by 2 as in (\ref{Sepsilon}) we find that the entanglement entropy is
\be
\label{Scft}
S_\epsilon = {c \over 12} L_\epsilon \sim {c \over 6} \log {1 \over \epsilon}
\ee
in agreement with  \cite{Calabrese:2004eu,Holzhey:1994we}.
An alternative perspective is to rescale by a factor $\pi/L_\epsilon$.  This keeps the size of the spatial interval fixed at $\pi$, but with an effective inverse temperature
\be
\beta_\epsilon = {\pi \over L_\epsilon} \beta \sim {\pi \beta \over 2 \log (1/\epsilon)}
\ee
Note that $\beta_\epsilon$ vanishes as $\epsilon \rightarrow 0$.  In this perspective entanglement entropy for a 2D CFT is simply the high-temperature limit of ordinary thermal entropy.

%%%%%%%%%%%%%%%%%%%%%%%%%%
\section{Massive scalar in 2D\label{sect:massive}}
%%%%%%%%%%%%%%%%%%%%%%%%%%
Next we consider a simple example of a non-conformal theory: a free massive scalar in two non-compact dimensions.  For this problem we set
\be
V_\epsilon = \int_{-\infty}^\infty dx \, r_\epsilon(x) T_{00}
\ee
and make the explicit choice $r_\epsilon(x) = \sqrt{x^2 + \epsilon^2}$.  The partition function $Z_\epsilon(\beta) = {\rm Tr} \, e^{-\beta V_\epsilon}$ corresponds to a Euclidean geometry
\be
ds^2 = dx^2 + \big(r_\epsilon(x)\big)^2 d\theta^2 \qquad -\infty < x < \infty,\quad \theta \approx \theta + \beta
\ee
Setting $x = \epsilon \sinh y$ makes the metric  conformally flat.
\be
\label{conformal}
ds^2 = \epsilon^2 \cosh^2y \left(dy^2 + d\theta^2\right) \qquad -\infty < y < \infty,\quad \theta \approx \theta + \beta
\ee
The action for a free massive scalar on this geometry is
\be
S = \int_{-\infty}^\infty dy \int_0^\beta d\theta \left({1 \over 2} (\partial_y \phi)^2 + {1 \over 2} (\partial_\theta \phi)^2 + {1 \over 2} m^2 \epsilon^2 \cosh^2 y \, \phi^2\right)
\ee
so that\footnote{In writing the partition function in this way we're making an implicit choice for the path integral measure \cite{Barbon:1994ej,Emparan:1994qa,deAlwis:1994ej},
one that corresponds to the optical geometry of \cite{Gibbons:1976pt}.}
\be
Z_\epsilon(\beta) = \det{}^{-1/2} \left(-\partial_y^2 - \partial_\theta^2 + m^2 \epsilon^2 \cosh^2 y\right)
\ee
Interpreting the $\theta$ direction as a thermal circle and passing to a Hamiltonian description we have a Bose partition function given by
\be
- \log Z_\epsilon = \sum_n \log \left(1 - e^{-\beta \omega_n}\right)
\ee
Here the frequencies $\omega_n$ are obtained from a Schrodinger problem with potential\footnote{We're denoting the Schrodinger potential ${\cal V}$ to distinguish it from the
operators $V$, $V_\epsilon$ we've introduced.} ${\cal V}(y) = m^2 \epsilon^2 \cosh^2 y$.
\be
\label{Schrodinger}
\left(-\partial_y^2 + m^2 \epsilon^2 \cosh^2 y\right) \psi_n = \omega_n^2 \psi_n
\ee
Classically the ground state sits at the minimum of the potential and has $\omega_0 = m\epsilon$.  For small $m \epsilon$ the partition function will be dominated
by highly excited states (recall that we're interested in $\beta \approx 2 \pi$).  To analyze this we follow 't Hooft \cite{tHooft:1984kcu} and approximate the spectrum as continuous, with
\bea
\nonumber
- \log Z_\epsilon & = & \int_0^\infty dn \, \log \left(1 - e^{-\beta \omega_n}\right) \\
\nonumber
& = & \int_{\omega_0}^\infty d\omega \, {dn \over d\omega} \log \left(1 - e^{-\beta \omega_n}\right) \\
\label{logZ}
& = & - \beta \int_{\omega_0}^\infty d\omega \, n(\omega) {1 \over e^{\beta \omega} - 1}
\eea
To get a rough approximation for $n(\omega)$, the number of states with frequency less than $\omega$, note that the Schrodinger problem has turning points at $y \approx \pm \log {2 \omega \over m \epsilon}$.  So we can approximate it
as a particle in a box with size $L = 2 \log {2 \omega \over m \epsilon}$.  For such a particle $\omega_n = {n \pi \over L}$ or
\be
\label{box}
n(\omega) = {L \omega \over \pi} = {2 \omega \over \pi} \log {2 \omega \over m \epsilon}
\ee
Using this in (\ref{logZ}) gives
\be
\label{ScalarLogZ}
- \log Z_\epsilon = - {\pi \over 3 \beta} \log {1 \over m \epsilon} + ({\rm finite})
\ee
Then the entanglement entropy (\ref{Sepsilon}) is
\be
S_\epsilon = {1 \over 6}  \log {1 \over m \epsilon} + ({\rm finite})
\ee
This simple approximation captures the leading log divergence in the entanglement entropy.  Note that the coefficient of the log is universal and agrees with (\ref{Scft}).  In appendix \ref{appendix:WKB} we re-derive these results from a more systematic WKB approximation.

%%%%%%%%%%%%%%%%%%%%%%%%%%
\section{Non-minimal scalar in 2D\label{sect:nonminimal}}
%%%%%%%%%%%%%%%%%%%%%%%%%%
In the examples we've considered so far the hourglass prescription reproduces the standard expressions for the entanglement entropy.  We now consider an example where the hourglass prescription makes a difference:
a scalar field in two dimensions with a non-minimal coupling to curvature.  The Euclidean action is
\be
S = \int d^2x \sqrt{g} \left({1 \over 2} g^{\mu\nu} \partial_\mu \phi \partial_\nu \phi + {1 \over 2} \xi R \phi^2 + {1 \over 2} m^2 \phi^2\right)
\ee
In the conformally flat coordinates introduced in (\ref{conformal}) the hourglass geometry is
\be
ds^2 = \epsilon^2 \cosh^2y \, \big(dy^2 + d\theta^2\big) \qquad -\infty < y < \infty,\quad \theta \approx \theta + \beta
\ee
with curvature
\be
\label{R}
R = - {2 \over \epsilon^2 \cosh^4 y}
\ee
Thus
\be
S = \int_{-\infty}^\infty dy \int_0^\beta d\theta \, \left({1 \over 2} \big(\partial_y \phi\big)^2 + {1 \over 2} \big(\partial_\theta \phi\big)^2 + {1 \over 2} {\cal V}(y) \phi^2\right)
\ee
where the effective potential
\be
\label{V}
{\cal V}(y) = m^2 \epsilon^2 \cosh^2 y - {2 \xi \over \cosh^2 y}
\ee
In a Hamiltonian description
\be
- \log Z_\epsilon = \sum_n \log \left(1 - e^{-\beta \omega_n}\right)
\ee
where the frequencies are determined by the Schrodinger equation
\be
\left(-\partial_y^2 + {\cal V}(y)\right) \psi_n = \omega_n^2 \psi_n
\ee

\begin{figure}
\centerline{\includegraphics[width=8cm]{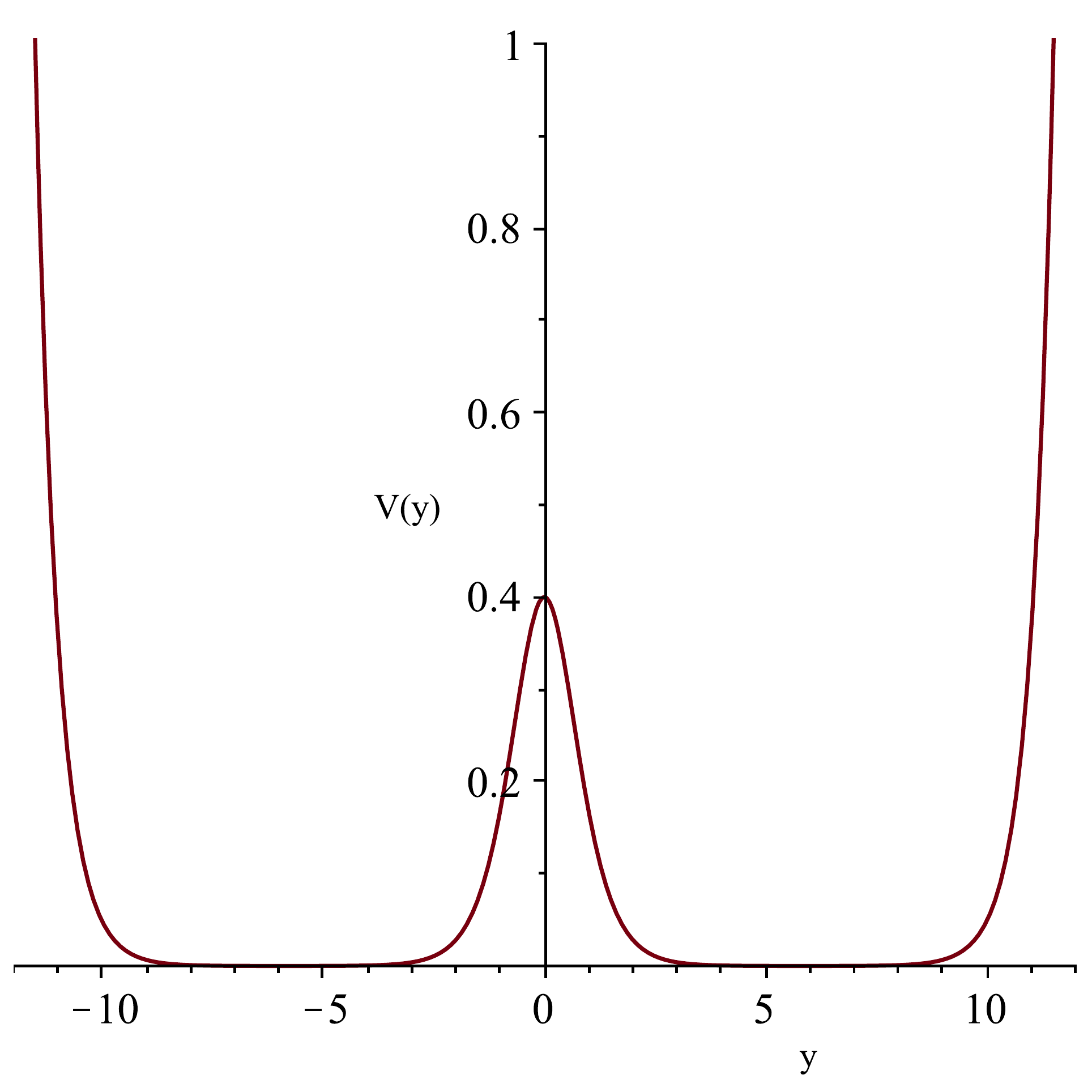}}
\caption{Effective potential for a non-minimal scalar with $m \epsilon = 2 \times 10^{-5}$ and $\xi = -0.2$.\label{fig:V}}
\end{figure}

From this description it's clear that the hourglass prescription gives a regulated entanglement entropy $S_\epsilon$ whose leading behavior as $\epsilon \rightarrow 0$ is insensitive to the non-minimal coupling.  That is, the hourglass
prescription gives $S_\epsilon = {1 \over 6} \log {1 \over m \epsilon} + {\rm (finite)}$ just as in section \ref{sect:massive}.  The reason is that the non-minimal coupling produces a bump or dip in the potential of height $\sim \xi$ and
width $\sim 1$ as shown in Fig.\ \ref{fig:V}.  But roughly speaking it's still a particle in a box of size $\sim 2 \log {2 \omega \over m \epsilon}$, and as $\epsilon \rightarrow 0$ the bump can be neglected relative to
the size of the box.  So by adopting the hourglass prescription for the entropy we find that the leading behavior of $n(\omega)$, and therefore the leading behavior of the entropy $S_\epsilon$, is insensitive to the
value of the non-minimal coupling parameter $\xi$.  We derive this result from a more systematic WKB approximation in appendix \ref{appendix:WKB}.

Of course this behavior was to be expected.  The non-minimal coupling modifies the stress tensor and controls how the theory couples to gravity.  But in flat space the non-minimal coupling has no effect on the
ground state wavefunction of the field and should have no effect on entanglement.  The hourglass prescription agrees with this expectation.  By contrast, the prescription of putting the theory on a cone and varying
with respect to the deficit angle leads to an entropy whose leading behavior depends on $\xi$: $S_{\rm cone} = \left({1 \over 6} - \xi\right) \log {1 \over m \epsilon}$ \cite{Solodukhin:1995ak}.

%%%%%%%%%%%%%%%%%%%%%%%%%%
\section{Maxwell field in 2D\label{sect:maxwell}}
%%%%%%%%%%%%%%%%%%%%%%%%%%
Next we explore the issues associated with gauge symmetry by considering a Maxwell field in two dimensions.  Since there are no local degrees of freedom we expect the entropy to vanish.  We will see that the hourglass prescription
agrees with this expectation.  We show this in two separate choices of gauge: first in a physical gauge, where the absence of local degrees of freedom makes the entropy vanish, then in a covariant gauge, where the
entropy vanishes due to a cancellation between gauge and ghost degrees of freedom.

%%%%%%%%%%%%%%%%%%%%%%%%%%
\subsection{Physical gauge\label{sect:weyl}}
%%%%%%%%%%%%%%%%%%%%%%%%%%
We first consider Maxwell theory in physical or unitary gauge.  We work in Lorentzian signature, in flat two-dimensional Minkowski space, and begin by writing the Maxwell action in first-order form.
\be
S = \int d^2x \, \left(-E_x \partial_t A_x - {1 \over 2} E_x^2 - A_t (\partial_x E_x - \rho)\right)
\ee
Here $(A_t,A_x)$ is the gauge field, $E_x$ is the electric field and $\rho$ is a specified charge density.  Integrating out $E_x$ using its equation of motion $E_x = \partial_x A_t - \partial_t A_x$
recovers the usual second-order action.  But in first-order form we see that $A_t$ is a Lagrange multiplier enforcing the Gauss constraint $\partial_x E_x = \rho$.

We'd like to make explicit that the hourglass prescription is consistent with having flux across the entangling surface.  To this end we introduce equal and opposite charges $\pm Q$ at fixed positions,
\be
\label{charges}
\rho = Q \delta(x + L) - Q \delta(x - L)
\ee
and solve the Gauss constraint by\footnote{More generally we could allow electric flux to extend to infinity.  We prefer to think of $L$ as an infrared cutoff and keep the flux confined to $-L < x < L$.}
\be
E_x = \left\lbrace\begin{array}{ll} Q & \qquad -L < x < L \\[3pt] 0 & \qquad {\rm otherwise} \end{array}\right.
\ee
The energy density is $T_{tt} = {1 \over 2} E_x^2$ so the operator $V$ defined in (\ref{Vdef}) is
\be
V = \int_{-L}^L dx \, \vert x \vert T_{tt} = {1 \over 2} L^2 Q^2
\ee
Note that, given the absence of local degrees of freedom, there is no need to introduce a UV regulator $\epsilon$.

Since there are no degrees of freedom there is nothing to quantize.  The partition function is rather trivially given by
\be
\label{ZLorentzian}
Z(\beta) = {\rm Tr} \, e^{-\beta V} = e^{-\beta L^2 Q^2 / 2}
\ee
As a result the entropy
\be
S = {1 \over 2} \left.\left(\beta {\partial \over \partial \beta} - 1\right)\right\vert_{\beta = 2 \pi} \big(- \log Z(\beta)\big)
\ee
vanishes.  It's no surprise that in a theory with no local degrees of freedom there is no entanglement.

%%%%%%%%%%%%%%%%%%%%%%%%%%
\subsection{Covariant gauge\label{sect:covariant}}
%%%%%%%%%%%%%%%%%%%%%%%%%%
For a different perspective, which will be useful below when we treat gauge fields in higher dimensions, we consider Maxwell theory in covariant gauge.
Of course we expect to obtain the same result, namely that the entropy vanishes.  But in covariant gauge this will be due to a rather non-trivial cancellation
between ghost and gauge degrees of freedom.  Our main challenge in this section will be understanding the boundary conditions that make this cancellation
possible.  Since the boundary conditions described here will not be important in the rest of the paper, the reader who is satisfied that entanglement entropy vanishes
in 2D Maxwell theory is invited to skip ahead to the next section.

We work on a two-dimensional Euclidean manifold with boundary.  Locally near the boundary we choose coordinates so that
$ds^2 = dn^2 + dt^2$ where ${\partial \over \partial n}$ is an inward-pointing normal vector and ${\partial \over \partial t}$ is tangent to the boundary.  The covariant gauge-fixed action is
\be
S_E = \int_{\cal M} d^2x \sqrt{g} \left({1 \over 4} F^2 + {1 \over 2} \left(\nabla_\mu A^\mu\right)^2 + i \partial_\mu b \partial^\mu c\right) + \int_{\partial {\cal M}} dt A_t J^t
\ee
We've included ghost fields $b,\,c$ and introduced a conserved current $J^t$ (really just a charge) on the boundary of the manifold.  We fix boundary conditions that are compatible with the boundary current we have introduced, namely
\bea
\nonumber
&& A_n \vert_{\partial {\cal M}} = 0 \\
\label{bcs}
&& F_{nt} \vert_{\partial {\cal M}} = \partial_n A_t \vert_{\partial {\cal M}} = J^t \\
\nonumber
&& \partial_n b \vert_{\partial {\cal M}} =  \partial_n c \vert_{\partial {\cal M}} = 0
\eea
In the literature these are variously known as absolute \cite{BransonGilkey}, electric \cite{Moss:1989wu} and type $I$ \cite{Vassilevich:1994we} boundary conditions.
To proceed we make a Hodge decomposition
\be
\label{Hodge}
A = d \phi_L + *d\phi_T + A_H
\ee
into longitudinal and transverse scalars $\phi_L,\, \phi_T$ plus a harmonic form $A_H$ that satisfies\footnote{Here $\delta = *d*$, not to be confused with the BRST variation $\delta_\alpha$ we introduce
below.  On a closed manifold a harmonic form would satisfy $d A_H = \delta A_H = 0$ but on a manifold with boundary (\ref{HarmonicCondition}) is the best we can do.}
\be
\label{HarmonicCondition}
(d \delta + \delta d)A_H = 0, \qquad \delta A_H = \nabla_\mu A_H^\mu = 0
\ee
The boundary conditions (\ref{bcs}) translate into
\bea
\label{bcphiL}
&& \partial_n\phi_L \vert_{\partial {\cal M}} = 0 \\
\label{bcphiT}
&& \phi_T \vert_{\partial {\cal M}} = 0 \\
\label{bcAHn}
&& A_{Hn} \vert_{\partial {\cal M}} = 0 \\
\label{bcAHt}
&& \partial_n A_{Ht} \vert_{\partial {\cal M}} = J^t \\
&& \partial_n b \vert_{\partial {\cal M}} =  \partial_n c \vert_{\partial {\cal M}} = 0
\eea
To see that these are the appropriate conditions to impose, note that according to the Hodge decomposition (\ref{Hodge}) the normal component of the gauge field $A_n$ and the field strength $F$ are given by
\bea
\nonumber
&&A_n = \partial_n \phi_L + \partial_t \phi_T + A_{Hn} \\
&&F = * \nabla^2 \phi_T + d A_H
\eea
On the boundary the conditions (\ref{bcphiL}), (\ref{bcphiT}), (\ref{bcAHn}) make the first line vanish while the conditions (\ref{bcphiT}), (\ref{bcAHn}), (\ref{bcAHt}) ensure that $F_{nt}$ from the second line is equal to $J^t$.  Thus the boundary conditions (\ref{bcs}) are satisfied.

It's important that the boundary conditions we have imposed respect the BRST symmetry ($\alpha$ is a Grassmann parameter)
\be
\delta_\alpha A_\mu = i \alpha \partial_\mu c \qquad \delta_\alpha b = \alpha \nabla_\mu A^\mu \qquad \delta_\alpha c = 0
\ee
To see how this works, note that under BRST
\bea
\nonumber
&&\delta_\alpha A_n = i \alpha \partial_n c \\
&&\delta_\alpha F = 0 \\
\nonumber
&&\delta_\alpha \partial_n b = \alpha \partial_n \nabla^2 \phi_L
\eea
On the boundary the first line vanishes by the Neumann condition on $c$, the second line trivially vanishes, and the third line vanishes since $\phi_L$ and therefore $\nabla^2 \phi_L$ obey Neumann
boundary conditions.  So the boundary conditions (\ref{bcs}) are preserved by BRST.

With boundary conditions in place let's analyze the one-loop determinants which arise in covariant gauge.  We will return to consider the harmonic piece below.  So leaving aside
$A_H$, under the Hodge decomposition (\ref{Hodge}) the vector wave operator becomes
\be
(d \delta + \delta d) A = d \nabla^2 \phi_L + * d \nabla^2 \phi_T
\ee
In place of $A_\mu$ we integrate over $\phi_L$ with Neumann boundary conditions and $\phi_T$ with Dirichlet boundary conditions.  The Neumann boundary conditions on $\phi_L$ allow a zero mode;
we suppress this zero mode on the grounds that a constant $\phi_L$ gives $A_\mu = 0$ and hence does not appear in the measure for integrating over $A_\mu$.  The ghosts obey Neumann
boundary conditions with zero modes which we likewise suppress.  Thus the one-loop determinants are
\be
\label{determinants}
\det{}^\prime{}^{-1/2}(-\nabla^2_N) \, \det{}^{-1/2}(-\nabla^2_D) \, \det{}^\prime{}(-\nabla^2_N) 
\ee
arising from $\phi_L$, $\phi_T$ and the ghosts, respectively.  The cancellation is not obvious unless the Dirichlet and Neumann Laplacians $\nabla^2_D$, $\nabla^2_N$ happen to be
(aside from the zero mode) isospectral.  Fortunately the hourglass geometry is conformal to a cylinder and on a cylinder the Dirichlet and Neumann Laplacians are isospectral.\footnote{Explicitly the Dirichlet and Neumann modes on a cylinder are related by $\sin \leftrightarrow \cos$.}
This makes it seem the determinants (\ref{determinants}) should cancel.  One should still worry about the conformal transformation from the hourglass to the cylinder.  But this is a local change in integration
measure which, as discussed in \cite{deAlwis:1994ej}, produces a Liouville action that doesn't affect the entropy.\footnote{Intuitively the change in integration measure, being local, shifts the effective action by a term
proportional to $\beta$ and hence does not affect the entropy.\label{MeasureFootnote}}  So for our purposes the fluctuation determinants cancel.

We still have to treat the harmonic part of the gauge field.  We do this explicitly on the hourglass geometry
\be
ds^2 = dx^2 + \big(r_\epsilon(x)\big)^2 d\theta^2 \qquad -L < x < L,\quad \theta \approx \theta + \beta
\ee
where we have introduced $L$ as an infrared cutoff.  As in (\ref{charges}) we place a charge $+Q$ on the left boundary and $-Q$ on the right boundary.  The harmonic gauge field is then
\bea
\nonumber
&& A_{Hx} = 0 \\
\label{AHtheta}
&& A_{H\theta} = {\rm const.} + Q \int_0^x dx' \, r_\epsilon(x')
\eea
with corresponding field strength
\be
F = dA_H = Q r_\epsilon(x) dx \wedge d\theta
\ee
This satisfies the harmonic condition $(d \delta + \delta d) A_H = 0$, $\delta A_H = 0$ or equivalently $\nabla_\mu F^{\mu \nu} = 0$, $\nabla_\mu A_H^\mu = 0$ as well as the boundary conditions
(\ref{bcAHn}), (\ref{bcAHt}).  The constant term in (\ref{AHtheta}) corresponds to a Polyakov loop that won't be important for our purposes.  The Euclidean action is
\be
S_E = {1 \over 2} \beta Q^2 \int_{-L}^L dx' r_\epsilon(x') + \beta Q^2 \int_0^{-L} dx' r_\epsilon(x') - \beta Q^2 \int_0^L dx' r_\epsilon(x')
\ee
arising from the bulk, left boundary and right boundary, respectively.  In total we have
\bea
\nonumber
S_E & = & - {1 \over 2} \beta Q^2 \int_{-L}^L dx' r_\epsilon(x') \\
& \rightarrow & - {1 \over 2} \beta Q^2 L^2 \qquad \hbox{\rm as $\epsilon \rightarrow 0$}
\eea
One final subtlety is that $Q$ here is the Euclidean charge which couples to $A_\theta$.  To relate it to the charge in Lorentzian signature we must Wick rotate $Q^2 \rightarrow - Q^2$.  After this
replacement we have
\be
Z(\beta) = e^{-S_E} = e^{-\beta Q^2 L^2 / 2}
\ee
in agreement with (\ref{ZLorentzian}).

%%%%%%%%%%%%%%%%%%%%%%%%%%
\section{Scalar fields in higher dimensions\label{sect:higher}}
%%%%%%%%%%%%%%%%%%%%%%%%%%
We now turn to theories in higher dimensions.  In this section we treat scalar fields with mass parameter $m$ and non-minimal coupling parameter $\xi$.  We introduce $d-2$ additional transverse
dimensions which we compactify on a torus of size $L_1 \times \cdots \times L_{d-2}$.  Our goal is to study the dependence of the entropy on all of these parameters.

To do this efficiently it's convenient to introduce a Schwinger proper-time representation for the partition function.  We will calculate the entropy while keeping in place a short-distance cutoff on proper time,
$s_{\rm min} = 1/\Lambda^2$.
This cutoff regulates UV divergences and will allow us to take the singular limit of the hourglass geometry, $\epsilon \rightarrow 0$, in which the hourglass reduces to a double cone.  (The singular limit can be
seen by reversing the arrow in Fig.~\ref{fig:hourglass}.)  The singular limit with a proper time cutoff provides a convenient way to package the dependence on the various parameters in the problem.  We should say
in advance that there are no physical surprises in the results.  For example we find that, as expected, the entropy has a leading power-law UV divergence proportional to the volume of the transverse torus.
But we do find an interesting mathematical connection, that the entropy is proportional to the late-time behavior of the heat kernel on an hourglass geometry.

We start from the Euclidean action
\be
S = \int d^dx \sqrt{g} \left({1 \over 2} g^{\mu\nu} \partial_\mu \phi \partial_\nu \phi + {1 \over 2} \xi R \phi^2 + {1 \over 2} m^2 \phi^2\right)
\ee
with metric
\bea
\label{HourglassTorus}
&& ds^2 = \underbrace{dx^2 + \big(r_\epsilon(x)\big)^2 d\theta^2}_{ds_2^2} + \underbrace{dz_1^2 + \cdots + dz_{d-2}^2}_{ds_{d-2}^2} \\[3pt]
\nonumber
&& -\infty < x < \infty,\quad \theta \approx \theta + \beta,\quad z_i \approx z_i + L_i
\eea
This describes a two-dimensional hourglass regulated by $\epsilon$, times a transverse torus of volume ${\rm vol}_\perp = L_1 \cdots L_{d-2}$.  An explicit choice is $r_\epsilon(x) = \sqrt{x^2 + \epsilon^2}$
but this is not essential to the results.

In a proper-time representation we have
\be
\label{SchwingerZ}
- \log Z = - {1 \over 2} \int_{1/\Lambda^2}^\infty {ds \over s} \, {\rm Tr} \, e^{-s(-\nabla^2 + \xi R + m^2)}
\ee
Here $\Lambda$ plays the role of a UV cutoff.  In short order this will allow us to simplify the geometry by sending $\epsilon \rightarrow 0$.  The mass acts as an IR cutoff.  Strictly speaking we need such a
cutoff because, although the torus cuts off IR divergences in the transverse directions, there are also IR divergences on the hourglass.

The heat kernel appearing in (\ref{SchwingerZ}) $K(s) = {\rm Tr} \, e^{-s(-\nabla^2 + \xi R + m^2)}$ can be decomposed as
\be
\label{factorization}
K(s) = K_2(s) K_{d-2}(s)
\ee
where
\bea
\nonumber
&& K_2(s) = {\rm Tr} \, e^{-s(-\nabla_2^2 + \xi R)} \qquad\, \hbox{\rm heat kernel on hourglass} \\
&& K_{d-2}(s) = {\rm Tr} \, e^{-s(-\nabla_{d-2}^2)} \qquad \hbox{\rm heat kernel on transverse torus}
\eea
The torus heat kernel is well known.  Its leading behavior at large volume (or small $s$) is
\be
K_{d-2}(s) = {{\rm vol}_\perp \over (4 \pi s)^{(d-2)/2}} \qquad \hbox{\rm at large volume}
\ee
This large-volume behavior is quite universal and would hold for any transverse geometry.  Here we focus on the large-volume behavior of the entropy.  We also focus on the leading UV divergence.
This means we can set $m = 0$ so that
\be
\label{leadingZ}
- \log Z = - {1 \over 2} \int_{1/\Lambda^2}^\infty {ds \over s} \, {{\rm vol}_\perp \over (4 \pi s)^{(d-2)/2}} \, K_2(s)
\ee
Corrections to this result, involving finite-mass and finite-volume effects, are analyzed in appendix \ref{appendix:corrections}.

The next challenge is to analyze $K_2(s)$.  This turns out to be surprisingly easy.  Setting $x = \epsilon \sinh y$ as in (\ref{conformal}) we have
\be
ds_2^2 = \epsilon^2 \cosh^2y \left(dy^2 + d\theta^2\right) = \epsilon^2 \widetilde{ds}_2^2
\ee
Note that $\epsilon$ only enters as an overall conformal factor, so the Laplacians are related by
\be
\nabla_2^2 = {1 \over \epsilon^2} \widetilde{\nabla}_2^2
\ee
and the heat kernels are related by
\be
\label{rescaling}
K_2(s) = \widetilde{K}_2(s/\epsilon^2)
\ee
Thus sending $\epsilon \rightarrow 0$ in (\ref{leadingZ}) gives
\be
\label{leadingZ2}
- \log Z = - {1 \over 2} \int_{1/\Lambda^2}^\infty {ds \over s} \, {{\rm vol}_\perp \over (4 \pi s)^{(d-2)/2}} \, \widetilde{K}_2(\infty)
\ee
We have the somewhat surprising result that $\log Z$ is proportional to the late-time behavior of the heat kernel on a unit hourglass ($\epsilon = 1$).

Although the short-time behavior is more familiar, we are able to obtain the late-time behavior of $\widetilde{K}_2(s)$ from the following observation.
The result (\ref{leadingZ2}) holds in any number of dimensions, in particular in $d = 2$.  Restoring the mass to serve as an IR cutoff, in two dimensions we have
\bea
- \log Z & = & - {1 \over 2} \int_{1/\Lambda^2}^\infty {ds \over s} \, e^{-sm^2} \, \widetilde{K}_2(\infty) \\
& = & \left(\log {m \over \Lambda} + {\rm finite}\right) \widetilde{K}_2(\infty)
\eea
Comparing the (universal) coefficient of the $\log$ divergence to our previous result (\ref{ScalarLogZ}) we identify\footnote{As in footnote \ref{MeasureFootnote}, this argument could miss terms arising from the
path integral measure that are proportional to $\beta$ and do not affect the entropy.}
\be
\label{LateTime}
\widetilde{K}_2(\infty) = {\pi \over 3 \beta}
\ee
It would be interesting to recover this result directly from an analysis of the spectrum of the hourglass Laplacian.
Note that, as shown in section \ref{sect:nonminimal}, this result is independent of the non-minimal coupling $\xi$.

With this in hand the rest of the calculation is immediate.  From (\ref{leadingZ2}) we have
\be
- \log Z = - {\pi \over 3 \beta} {{\rm vol}_\perp \Lambda^{d-2} \over (d-2) (4\pi)^{(d-2)/2}}
\ee
and
\bea
\nonumber
S_{\rm scalar} & = & {1 \over 2} \left.\left(\beta {\partial \over \partial \beta} - 1\right)\right\vert_{\beta = 2 \pi} \big(- \log Z(\beta)\big) \\
\label{Sscalar}
& = & {1 \over 6} {{\rm vol}_\perp \, \Lambda^{d-2} \over (d-2) (4\pi)^{(d-2)/2}}
\eea

%%%%%%%%%%%%%%%%%%%%%%%%%%
\section{Maxwell in higher dimensions\label{sect:higherMaxwell}}
%%%%%%%%%%%%%%%%%%%%%%%%%%
Finally we consider a Maxwell field in higher dimensions.  In covariant gauge the Euclidean action is
\be
S = \int d^dx \sqrt{g} \left({1 \over 4} F_{\mu\nu}F^{\mu\nu} + {1 \over 2} \big(\nabla_\mu A^\mu\big)^2 + i \partial_\mu b \partial^\mu c\right)
\ee
As in the scalar case we work on the geometry (\ref{HourglassTorus}) which is a product of an hourglass and a transverse torus.
\bea
&& ds^2 = \underbrace{dx^2 + \big(r_\epsilon(x)\big)^2 d\theta^2}_{ds_2^2} + \underbrace{dz_1^2 + \cdots + dz_{d-2}^2}_{ds_{d-2}^2} \\[3pt]
\nonumber
&& -\infty < x < \infty,\quad \theta \approx \theta + \beta,\quad z_i \approx z_i + L_i
\eea
It's convenient to break $x^\mu$ up into coordinates $x^\alpha = (x,\theta)$ on the hourglass and coordinates $x^i = z^i$ on the torus.  The gauge field has polarizations $A_\alpha$ tangent to the hourglass
and polarizations $A_i$ tangent to the torus.  Integrating by parts the action decomposes as
\be
S = \int d^dx \sqrt{g} \left[{1 \over 2} A^\alpha \left(-g_{\alpha\beta} \nabla_\mu \nabla^\mu + R_{\alpha\beta}\right)A^\beta + {1 \over 2} A^i \left(-\delta_{ij}\nabla_\mu\nabla^\mu\right)A^j - i b \nabla_\mu \nabla^\mu c\right]
\ee
The partition function is then
\bea
\nonumber
- \log Z & = & - {1 \over 2} \int_{1/\Lambda^2}^\infty {ds \over s} \, \left({\rm Tr} \, e^{-s(-g_{\alpha\beta} \nabla_\mu\nabla^\mu + R_{\alpha\beta})} + {\rm Tr} \, e^{-s(-\delta_{ij} \nabla_\mu\nabla^\mu)}\right) \\[5pt]
& & +  \int_{1/\Lambda^2}^\infty {ds \over s} \, {\rm Tr} \, e^{-s(-\nabla_\mu\nabla^\mu)}
\eea
As in (\ref{factorization}) the heat kernels in this expression factorize, and the torus polarizations make the same contribution as $d-2$ scalars.  Thus
\bea
\nonumber
- \log Z & = & - {1 \over 2} \int_{1/\Lambda^2}^\infty {ds \over s} \, \left(K_2^{\rm vector}(s)K_{d-2}^{\rm scalar}(s) + (d-2) K_d^{\rm scalar}(s)\right) \\[5pt]
& & +  \int_{1/\Lambda^2}^\infty {ds \over s} \, K_2^{\rm scalar}(s)K_{d-2}^{\rm scalar}(s)
\eea
In section \ref{sect:covariant} we showed that the Maxwell entropy vanishes in two dimensions.  This means that as far as the entropy is concerned
the first and third terms cancel and we're left with the entropy of $d-2$ scalar fields.  From (\ref{Sscalar}) we have
\be
S_{\rm Maxwell} = {1 \over 6} {{\rm vol}_\perp \, \Lambda^{d-2} \over (4\pi)^{(d-2)/2}}
\ee

%%%%%%%%%%%%%%%%%%%%%%%%%%
\section{Conclusions\label{sect:conclusions}}
%%%%%%%%%%%%%%%%%%%%%%%%%%
In this paper we have explored a prescription for defining entanglement entropy in quantum field theory.  There are two broad perspectives one can take on the prescription.

From a canonical or thermal perspective, the prescription avoids any factoring of the Hilbert space.  Instead the entropy is defined in terms
of the full Hilbert space and has a straightforward state-counting interpretation, as half of the von Neumann entropy of the density matrix (\ref{rho-epsilon}).
\be
\rho_\epsilon = {1 \over Z_\epsilon(\beta)} e^{-\beta V_\epsilon}
\ee
This density matrix has a thermal interpretation, corresponding to the position-dependent proper temperature (\ref{Tproper}).
\be
T_{\rm proper}(x) = {1 \over \beta r_\epsilon(x)}
\ee

From a covariant or path integral perspective the entropy can be obtained from the partition function on the product of an hourglass geometry with a transverse space.  This provides a powerful calculational
tool which is most effective when taking the singular limit of the hourglass geometry $\epsilon \rightarrow 0$ with a proper time cutoff in place.  Rather curiously the overall coefficient in the entropy is related
to the late-time behavior of the heat kernel on a unit hourglass.

In relating these two perspectives it's important that (just like an ordinary thermal system) the hourglass has a freely-acting Killing vector $\partial \over \partial \theta$.  This means local geometric quantities such as the curvature will be independent of $\theta$ and local terms in the effective action
will be proportional to $\beta$.  As in footnote \ref{MeasureFootnote}, such terms can shift the energy $E = {\partial \over \partial \beta} \big(-\log Z\big)$ but don't contribute to the entropy.  Instead, just as in an ordinary thermal system, only worldlines which wind around the Euclidean
time direction contribute to the entropy.  This stands in contrast to the prescription mentioned in the introduction, of putting the field theory on a cone and varying with respect to the deficit angle.  The Killing vector on a cone
has a fixed point.  This means local terms in the effective action can, and in the case of the Einstein-Hilbert term do, contribute to the entropy \cite{Susskind:1994sm}.

To conclude, let us mention a few directions for future work.
\begin{itemize}
\item
The leading divergence of the entropy can be thought of as counting degrees of freedom per unit area.  For example for a scalar field from (\ref{Sscalar}) we have
\be
{S \over {\rm vol}_\perp} = {\Lambda^{d-2} \over 6 (d-2) (4\pi)^{(d-2)/2}}
\ee
We expect that this quantity should be positive in any unitary theory.  Intuitively it should decrease under RG flow.  Does it satisfy a c-theorem?
\item
We have restricted our attention to planar entangling surfaces.  It would be interesting to consider other possibilities.
For a generic entangling surface in even dimensions it's reasonable to expect a log divergent term in the entanglement entropy.  It would be interesting to calculate
the coefficient of the log for a spherical entangling surface in 3+1 and compare to the results of \cite{Soni:2016ogt}.
\item
In a holographic setting one could in principle use the hourglass prescription to calculate entanglement in the boundary theory.  Can one identify the bulk dual of this calculation, say along the lines of \cite{Lewkowycz:2013nqa}?
\end{itemize}

%%%%%%%%%%%%%%%%%%%
\bigskip
\goodbreak
\centerline{\bf Acknowledgements}
\noindent
DK is grateful to Edward Witten for valuable discussions at the beginning of this work.
The work of TA and NI were supported in part by JSPS KAKENHI Grant Number 21J20906(TA), 18K03619(NI). 
The work of NI was also supported by MEXT KAKENHI Grant-in-Aid for Transformative Research Areas A ``Extreme Universe'' No.\ 21H05184.
DK is supported by U.S.\ National Science Foundation grant PHY-2112548.

\appendix
%%%%%%%%%%%%%%%%%%%%%%%%%%
\section{WKB for scalar fields\label{appendix:WKB}}
%%%%%%%%%%%%%%%%%%%%%%%%%%
We first consider a minimally-coupled scalar field.  The WKB quantization condition for the Schrodinger problem (\ref{Schrodinger}) is
\be
2 \int_0^{\cosh^{-1}(\omega/m\epsilon)} dy \, \sqrt{\omega^2 - m^2 \epsilon^2 \cosh^2 y} = \left(n + {1 \over 2}\right) \pi
\ee
Setting $u = {m \epsilon \over \omega} \cosh y$, $a = {m \epsilon \over \omega}$ this becomes
\be
n + {1 \over 2} = {2 \omega \over \pi} \int_a^1 du \, \sqrt{1 - u^2 \over u^2 - a^2}
\ee
The integral can be evaluated as the difference of two complete elliptic integrals and can be expanded for small $a$.
\bea
\nonumber
\int_a^1 du \, \sqrt{1 - u^2 \over u^2 - a^2} & = & K\big(\sqrt{1-a^2}\big) - E\big(\sqrt{1-a^2}\big) \\
\label{minimalI}
& = & \log {1 \over a} + \log 4 - 1 + {\cal O}(a^2)
\eea
Thus the leading behavior of $n(\omega)$ for small $a$ is
\be
\label{n}
n(\omega) = {2 \omega \over \pi} \log {\omega \over m \epsilon}
\ee
The same leading behavior was obtained from the particle-in-a-box approximation in (\ref{box}).

Next we consider scalar fields with non-minimal couplings.  For the potential (\ref{V}) the WKB quantization condition becomes
\be
\label{non-min}
n + {1 \over 2} = {2 \omega \over \pi} \int_{{\rm max}(a,u_-)}^{u_+} {du \over u} \sqrt{(u_+^2 - u^2) (u^2 - u_-^2) \over u^2 - a^2}
\ee
The variables $u$, $a$ are defined as above and the turning points are at
\be
u_\pm^2 = {1 \over 2} \pm {1 \over 2} \sqrt{1 - {8 \xi a^2 \over \omega^2}}
\ee
($u_-$ is only a turning point if $u_- > a$).  Note that when $a$ is small $u_- \approx {\sqrt{2 \xi} \over \omega} \, a$.

To evaluate the integral in (\ref{non-min}) for small $a$ with $u_- \sim a$ we introduce a parameter $z$ with $a \ll z \ll u_+$.  (A systematic choice is $z \sim \sqrt{a}$.)
We break the integral into two regions, $I = I_1 + I_2$ where
\bea
\nonumber
&& I_1 = \int_{{\rm max}(a,u_-)}^{z} {du \over u} \sqrt{(u_+^2 - u^2) (u^2 - u_-^2) \over u^2 - a^2} \\
&& I_2 = \int_z^{u_+} {du \over u} \sqrt{(u_+^2 - u^2) (u^2 - u_-^2) \over u^2 - a^2}
\eea
For small $a$ we can approximate
\bea
\nonumber
&& I_1 \approx u_+ \int_{{\rm max}(a,u_-)}^{z} {du \over u} \sqrt{u^2 - u_-^2 \over u^2 - a^2} \\[5pt]
&& I_2 \approx \int_z^{u_+} {du \over u} \sqrt{u_+^2 - u^2}
\eea
These integrals can be evaluated to obtain\footnote{There are two possible lower limits, $a$ and $u_-$, but the result (\ref{I}) is correct for both possibilities.}
\bea
\nonumber
I &=& \log {1 \over a} + \log 4 - 1 - {1 \over 4} \left(1 - {\sqrt{2 \xi} \over \omega}\right) \log \left(1 - {\sqrt{2 \xi} \over \omega}\right)^2 - {1 \over 4} \left(1 + {\sqrt{2 \xi} \over \omega}\right) \log \left(1 + {\sqrt{2 \xi} \over \omega}\right)^2 \\[5pt]
\label{I}& & + \, (\hbox{\rm terms which vanish as $a \rightarrow 0$})
\eea
Note that the non-minimal coupling only appears in the finite terms.  So for a non-minimal scalar the leading behavior of $n(\omega)$ is still given by (\ref{n}).

%%%%%%%%%%%%%%%%%%%%%%%%%%
\section{Finite mass and volume effects\label{appendix:corrections}}
%%%%%%%%%%%%%%%%%%%%%%%%%%
In this appendix we revisit the expression for a scalar field given in (\ref{SchwingerZ}).
\be
- \log Z = - {1 \over 2} \int_{1/\Lambda^2}^\infty {ds \over s} \, {\rm Tr} \, e^{-s(-\nabla^2 + \xi R + m^2)}
\ee
The leading UV divergent contribution to the entropy, given in (\ref{Sscalar}), is extensive in the volume of the transverse torus and is insensitive to the mass.
Our goal here is to understand corrections to this leading behavior arising from finite-mass and finite-volume effects.

Thanks to the factorization (\ref{factorization}), the rescaling (\ref{rescaling}) and the late-time behavior (\ref{LateTime}) we have\footnote{Note that the late-time behavior is completely insensitive
to the non-minimal coupling parameter $\xi$.}
\be
- \log Z = - {\pi \over 6 \beta} \int_{1/\Lambda^2}^\infty {ds \over s} \, e^{-sm^2} \, K_{d-2}(s)
\ee
The heat kernel for the transverse torus is
\be
K_{d-2}(s) = K(s,L_1) \cdots K(s,L_{d-2})
\ee
where the heat kernel for a circle (a periodic dimension of size $L$) can be written as a sum over momentum modes.
\be
\label{momentum}
K(s,L) = \sum_{n = - \infty}^\infty e^{-s(2\pi n / L)^2}
\ee
By Poisson resummation this can be written as a sum over winding modes.
\be
\label{winding}
K(s,L) = {L \over \sqrt{4 \pi s}} \sum_{n = - \infty}^\infty e^{-L^2 n^2 / 4 s}
\ee
Here $n$ is the number of times the particle worldline wraps around the circle.  This winding form will be more useful for our purposes.

We begin by considering corrections due to finite mass.  To do this we keep the UV cutoff $\Lambda$ in place and hold the mass fixed but take the torus to be large, $L_i \rightarrow \infty$ for all $i$.
Then only the zero-winding modes in (\ref{winding}) contribute and we have
\be
\label{FiniteMass}
- \log Z = - {\pi \over 6 \beta} \, {{\rm vol}_\perp \over (4 \pi)^{(d-2)/2}} \, I_d
\ee
where
\be
I_d = \int_{1/\Lambda^2}^\infty {ds \over s^{d/2}} \, e^{-sm^2}
\ee
It's straightforward to expand $I_d$ in powers of $m / \Lambda$, as illustrated in table \ref{table:Id}.  The entropy ${1 \over 2} \left(\beta \partial_\beta - 1\right)\vert_{\beta = 2\pi} \left(- \log Z\right)$ is given by
\be
S = {1 \over 12} \, {{\rm vol}_\perp \over (4 \pi)^{(d-2)/2}} \, I_d
\ee
As expected the entropy is extensive in the volume of the torus, with a leading divergence $\sim \Lambda^{d-2}$ and a series of corrections suppressed by powers of $m / \Lambda$.

\begin{table}
\begin{center}
\begin{tabular}{l|l}
$d$ \quad & \qquad\quad $I_d$ \\
\hline
2 & \quad $2 \log {\Lambda \over m} - \gamma \phantom{\Bigg]} $ \\[10pt]
3 & \quad $2 \Lambda - 2 \sqrt{\pi} m  $ \\[10pt]
4 & \quad $\Lambda^2 - 2 m^2 \log {\Lambda \over m} + (\gamma - 1) m^2 $ \\[10pt]
5 & \quad ${2 \over 3} \Lambda^3 - 2 \Lambda m^2 + {4 \over 3} \sqrt{\pi} m^3 $ \\[10pt]
6 & \quad ${1 \over 2} \Lambda^4 - \Lambda^2 m^2 + m^4 \log {\Lambda \over m} + \left({3 \over 4} - {\gamma \over 2}\right) m^4 $
\end{tabular}
\end{center}
\caption{$I_d$ in various dimensions, neglecting terms which vanish as $\Lambda \rightarrow \infty$.\label{table:Id}}
\end{table}

Now let's consider corrections due to finite-volume effects.  That is, we consider terms in (\ref{winding}) with non-zero winding.  Since these terms are exponentially suppressed at small $s$
they are UV finite which means we can send $\Lambda \rightarrow \infty$.\footnote{The various limits are a bit subtle.  To study finite mass effects we kept $m$ and $\Lambda$ fixed and sent
$L_i \rightarrow \infty$.  To study finite volume effects we instead keep $m$ and $L_i$ fixed and send $\Lambda \rightarrow \infty$.}  For concreteness we set $d=3$, meaning a single circle of size $L$,
and find the winding contribution
\be
- \log Z_{\rm winding} = - {\pi \over 6 \beta} \int_0^\infty {ds \over s} \, e^{-sm^2} \, {L \over \sqrt{\pi s}} \sum_{n = 1}^\infty e^{-n^2 L^2 / 4 s}
\ee
The integral and sum are straightforward.  Overall we find
\be
- \log Z_{d = 3} = \big(\hbox{\rm extensive result from (\ref{FiniteMass})}\big) + {\pi \over 3 \beta} \log \left(1 - e^{-Lm}\right)
\ee
As expected the winding contributions are exponentially suppressed at large $Lm$.  At small $Lm$ there is a divergence $\sim \log Lm$.  This is nothing but the IR divergence associated with the
hourglass.  The same divergence is present in $d = 2$ as can be seen in (\ref{ScalarLogZ}).  To make this connection precise note that the IR divergence arises from the $n = 0$ term in the momentum
sum (\ref{momentum}), and that setting $n = 0$ in that sum is equivalent to dimensional reduction to $d = 2$.

%\bibliographystyle{utphys}
%\bibliography{entropy}
\providecommand{\href}[2]{#2}\begingroup\raggedright\endgroup

\end{document}